\documentclass[twocolumn,prbrc]{revtex4}% Physical Review B RC

\usepackage{graphicx}
\usepackage{dcolumn}
\usepackage{amsmath}

\begin{document}
\preprint{HEP/123-qed}

\title[Short Title]{Cyclotron Resonance in Ferromagnetic InMnAs/(Al,Ga)Sb
Heterostructures}

\author{G. A. Khodaparast}
\author{J. Kono}
 \thanks{Author to whom correspondence should be addressed}
 %\homepage{http://www.ece.rice.edu/~kono}
 \email{kono@rice.edu}
\affiliation{Department of Electrical and Computer Engineering,
Rice Quantum Institute, and Center for Nanoscale Science and Technology,
Rice University, Houston, Texas 77005}
\author{Y. H. Matsuda}
\thanks{Present address: Department of Physics, Faculty of Science,
Okayama University, Okayama 700-8530, Japan.}
%\author{T. Ikaida}
\author{S. Ikeda}
\author{N. Miura}
\affiliation{Institute for Solid State Physics, University of Tokyo,
Kashiwa, Chiba 277-8581, Japan}
\author{T. Slupinski}
\thanks{Present address: Institute of Experimental Physics,
Warsaw University, Hoza 69, 00-681 Warsaw, Poland.}
\author{A. Oiwa}
\thanks{Present address: PRESTO, Japan Science and Technology
Corporation, 4-1-8 Honcho, Kawaguchi, Saitama 332-0012, Japan.}
\author{H. Munekata}
\affiliation{Imaging Science and Engineering Laboratory, Tokyo
Institute of Technology, Yokohama, Kanazawa 226-8503,
and Kanazawa Academy of Science and Technology, Kawasaki,
Kanazawa 213-0012, Japan}

\author{Y. Sun}
\author{F. V. Kyrychenko}
\author{G. D. Sanders}
\author{C. J. Stanton}
\affiliation{Department of Physics, University of Florida, Box 118440,
Gainesville, Florida 32611-8440}

\date{\today}

\begin{abstract}
We report the observation of hole cyclotron
resonance (CR) in InMnAs/(Al,Ga)Sb heterostructures in a wide temperature
range covering both the paramagnetic and ferromagnetic phases.
We observed two pronounced resonances that exhibit drastic changes in position,
linewidth, and intensity at a temperature higher than the Curie temperature,
indicating possible local magnetic ordering or clustering.
%One resonance abruptly reduces its linewidth, shifts its resonance
%field, and increases its intensity around $T_c$.
%The other resonance, which is absent at room temperature,
%suddenly appeares above $T_c$, rapidly grows in intensity with
%decreasing temperature, and becomes comparable to the first
%resonance at low temperatures.
We attribute the two resonances to the fundamental CR transitions
expected for delocalized valence-band holes in the
quantum limt.  Using an 8-band {\bf k$\cdot$p} model,
which incorporates ferromagnetism within a mean-field approximation,
we show that the temperature-dependent CR peak shift is a direct measure
of the carrier-Mn exchange interaction.
Significant line narrowing was observed at low temperatures, which we
interpret as the suppression of localized spin fluctuations.

\end{abstract}
\maketitle

%\tableofcontents
%%%%%%%%%%%%%%%%%%%%%%%%%%%%%%%%%%%%%%%%%%%%%%%%%%%%%%%%%%%%%%%%%%%%%%%%%%%%%%%%%%%%%%%%%%%%%%%%%%%%%%%%%%%
%Introduction
The interaction of free carriers with localized spins plays an
important role in a variety of physical situations in metals
\cite{anderson,kittel,hewson}.  Carriers near a magnetic ion are
spin-polarized, which can mediate an
indirect exchange interaction between magnetic ions.  The carrier-induced
ferromagnetism realized in Mn-doped III-V semiconductors \cite{ohno}
has provided a novel system in which to
study itinerant carriers interacting with
localized spins in the dilute limit.  Various mechanisms have been suggested
but the microscopic origin of carrier-induced
ferromagnetism is still controversial \cite{theory}.
One of the unanswered questions is the nature of the carriers, i.e.,
whether they
are in the impurity band ($d$-like), the delocalized
valence bands ($p$-like), or some type of mixed states.

Here we report a clear observation of hole cyclotron
resonance (CR) in ferromagnetic InMnAs/(Al,Ga)Sb,
demonstrating the existence of delocalized $p$-like
carriers.  In addition, this is the first CR study
in any ferromagnet covering temperatures ($T$'s)
both below and above the Curie temperature ($T_c$) \cite{grimes}.
CR is a direct method for determining carrier effective masses and
therefore the nature of the carrier states.  In all samples
studied, we observed two resonances, both of which
exhibited unusual $T$-dependence in position,
intensity, and width.
The lower-field line showed an abrupt reduction in
width with a concomitant decrease in resonance field
above $T_c$.  The higher-field line, which was absent
at room $T$, suddenly appeared above $T_c$, rapidly
grew with decreasing $T$, and became
comparable to the lower-field resonance at low $T$.
We ascribe these lines to the two fundamental CR transitions
expected for delocalized holes in the valence band of a
Zinc-Blende semiconductor in the magnetic quantum limt.
We developed an 8-band {\bf k$\cdot$p} theory, including full valence
band complexities and treating ferromagnetism
within a mean-field approximation.  Results show that the $T$-dependent
CR peak shift
is a direct measure of the carrier-Mn exchange interaction.
%We provide qualitative explanations to the unusual
%temperature dependence based on the expected band structure
%changes due to ferromagnetic order.
%We attempted similar measurements
%on GaMnAs but did not detect any sign of resonance, which
%indicates that the nature of carriers in
%these two systems (InMnAs and GaMnAs) may be very different.
%Our experiment and theory
%provide significant new information on the carrier states
%in this family of ferromagnetic semiconductors.

\begin{table}[bth]
\caption{Characteristics of the In$_{1-x}$Mn$_x$As/Al$_y$Ga$_{1-y}$Sb
samples.
Densities and mobilities are room $T$ values.
$m_A$ and $m_B$ are the low-$T$ cyclotron masses for the two lines
(see Fig. \ref{typical}).}
\label{table1}
\begin{ruledtabular}
\begin{tabular*}{\hsize}{l@{\extracolsep{0ptplus1fil}}c@{\extracolsep{
0ptplus1fil}}c@{\extracolsep{0ptplus1fil}}c@{\extracolsep{0ptplus1fil}}c}
Sample No. & 1 & 2 & 3 & 4\\
\colrule
$T_c$ (K) & 55 & 30 & 40 & 35\\
Mn content $x$ & 0.09 & 0.12 & 0.09 & 0.12\\
Al content $y$ & 0 & 0 & 0 & 1\\
Thickness (nm) & 25 & 9 & 31 & 9\\
%$d$(GaSb) (nm) & 820 & 600 & 800 \\
Density (cm$^{-3}$) & 1.1$\times$10$^{19}$ & 4.8$\times$10$^{19}$ &
1.1$\times$10$^{19}$ & 4.8$\times$10$^{19}$\\
Mobility (cm$^2$/Vs) & 323 & 371 & 317 & 384\\
$m_A$/$m_0$ & 0.0508 & 0.0525 & 0.0515 & 0.0520\\
$m_B$/$m_0$ & 0.122 & 0.125 & 0.125 & 0.127\\
\end{tabular*}
\end{ruledtabular}
\end{table}

The samples were In$_{1-x}$Mn$_x$As/Al$_y$Ga$_{1-y}$Sb
single heterostructures containing high densities ($\sim$10$^{19}$
cm$^{-3}$) of holes.
They were grown by low temperature molecular beam epitaxy
on GaAs (100) substrates \cite{tom}.  Unlike the $n$- and $p$-type
films we studied earlier \cite{matsuda,zudov,matsuda1,sanders_jap},
the samples in the present work showed ferromagnetism.
%The magnetization easy axis
%was perpendicular to the epilayers due to the strain-induced
%structural anisotropy.
Sample characteristics are summarized in Table I.
%The
%densities and mobilities were estimated from room temperature Hall
%measurements and $T_c$'s were determined by
%magnetization measurements.
Sample 1 was annealed at
250$^{\circ}$C after growth, which increased the $T_c$
by $\sim$10 K \cite{hayashi}.
%Finally, the low temperature cyclotron masses deduced from
%CR discussed below are also listed in Table I.
We performed CR measurements using ultrahigh pulsed
magnetic fields ($B$'s) generated by the single-turn coil technique
\cite{zudov,nakao}.
%The field ($B$) was measured
%by a pick-up coil around the sample, which was placed inside a
%continuous flow helium cryostat.
We used circularly-polarized radiation with wavelengths of 10.6 $\mu$m,
10.2 $\mu$m, 9.25 $\mu$m (CO$_2$ laser), and 5.527 $\mu$m (CO laser), and
the transmitted radiation was detected using a
fast HgCdTe detector.
A multi-channel digitizer recorded the
signals from the pick-up coil and the detector.
%Although the coil
%breaks in each shot, the sample and pick-up coil remain
%intact, making it possible to
%carry out detailed temperature and wavelength dependence studies
%on the same specimen.
%Since the
%transmission signal was recorded during both the up and down
%sweeps, each resonance was observed twice in a single pulse.
%This allowed us to check the reproducibility of observed
%absorption peaks and to make sure that the spectra were free
%from any slow heating effects.

%%%%%%%%%%%%%%%%%%%%%%%%%%%%%%%%%%%%%%%%%%%%%%
\begin{figure}
\begin{center}
%\resizebox{0.5\textwidth}{!}{
 \includegraphics[scale=0.9]{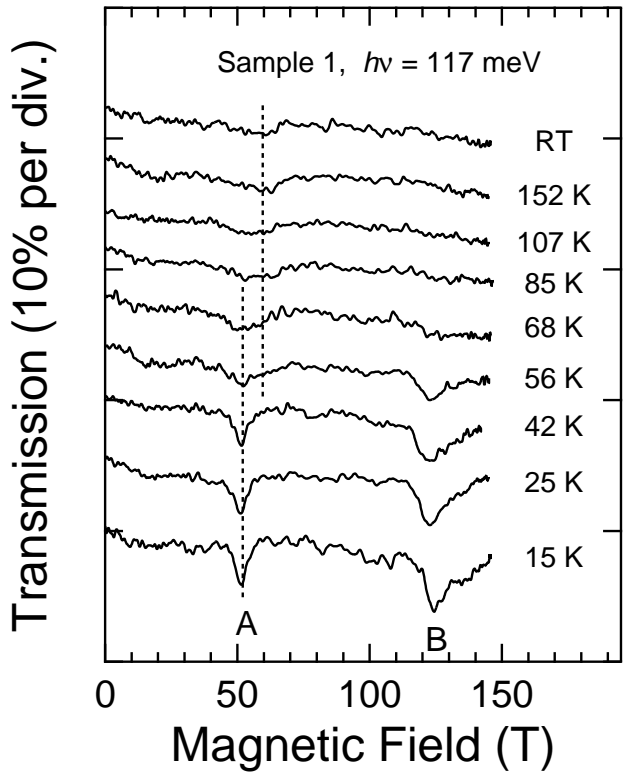}
%}
\caption{CR spectra for sample 1.  The transmission of hole-active circular
polarized 10.6 $\mu$m radiation
is plotted vs. magnetic field at different temperatures.
%Both samples show two strongly temperature-dependent features,
%labeled A and B, whose origins are discussed in the text.
}
\label{typical}
\end{center}
\end{figure}
%%%%%%%%%%%%%%%%%%%%%%%%%%%%%%%%%%%%%%%%%%%%%%%

Figure \ref{typical} shows the transmission of the 10.6
$\mu$m beam through sample 1 at various $T$'s as a function of $B$.
From room $T$ down to slightly above
$T_c$, a broad feature (labeled `A') is
observed with almost no change in intensity,
position, and width with decreasing $T$.  However,
at $\sim$68 K, which is sill above $T_c$, we observe quite abrupt
and dramatic changes in the spectra.  First, a significant
reduction in linewidth and a sudden shift to a
lower $B$ occur simultaneously.  Also, it
increases in intensity rapidly with decreasing $T$.
In addition, a second feature (labeled `B') suddenly appears
$\sim$ 125 T, which also rapidly grows in intensity with
decreasing $T$ and saturates, similar to feature A.
%Essentially the same
%behavior is seen for sample 2 in Fig. 1(b).
%Note that in both cases {\em the temperature at which
%the unusual spectral changes occur is higher
%than $T_c$.}
%From Lorentzian fits to the data, we deduced the
%cyclotron masses, densities, and mobilities.
%The cyclotron densities and mobilities deduced from Lorentzian fits to the
%data for feature A are plotted in Fig. 2(a), together with the $T$-dependence of
%the magnetization $M$ in Fig. 2(b).
%It is clear from these plots that the two quantities deduced from
Note that {\em the
temperature at which these unusual sudden CR changes occur ($T_c^*$) is higher
than $T_c$.}

\begin{figure}
\begin{center}
%\resizebox{0.5\textwidth}{!}{
 \includegraphics[scale=1.0]{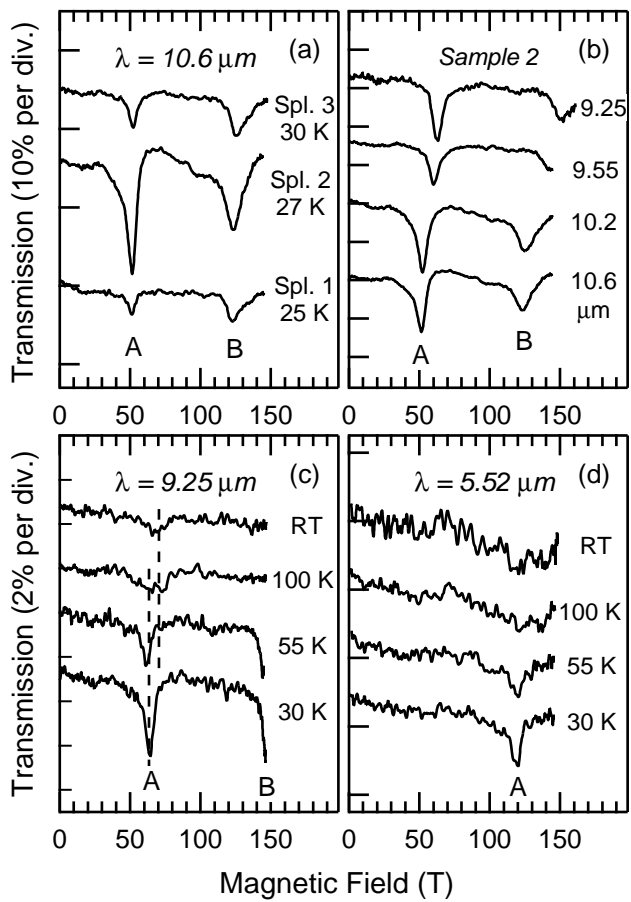}
%}
\caption{
(a) Low temperature CR spectra for three samples at
10.6 $\mu$m.  (b) Wavelength dependence of the CR spectra for
sample 2 at 27 K.  CR speatra for sample 1
at different temperatures at (c) 9.25 $\mu$m and (d) 5.52 $\mu$m.
}
\label{various}
\end{center}
\end{figure}

The observed unusual $T$-dependence is neither specific to this
particular wavelength ($\lambda$) used nor to the sample measured.
We observed essentially the same $T$-dependent behavior in all the
samples studied.  Figure \ref{various}(a) shows low-$T$
CR traces for three samples at 10.6 $\mu$m.
Both features A and B are clearly
observed but their intensities and linewidths vary from sample
to sample.
Figure \ref{various}(b) displays the $\lambda$-dependence of the CR
spectra for sample 2.  We can see that both lines shift
to higher $B$'s with decreasing $\lambda$ (i.e.,
increasing photon energy), as expected.
Figures \ref{various}(c) and \ref{various}(d) show data at different $T$'s
for sample 1 measured at 9.25 $\mu$m and 5.52 $\mu$m, respectively.
The $T$-dependence observed
at these shorter $\lambda$'s is
similar to what was observed at 10.6 $\mu$m.
The observations of CR
with essentially the same masses in samples with different buffer
layers (GaSb or AlSb)
exclude the possibility of hole CR in the buffer.  We also confirmed
the absence of CR in a control sample which consisted of only a GaSb layer
grown on GaAs.  All these facts confirm the universality of the effects we
observed and their relevance to ferromagnetic order.

%%%%%%%%%%%%%%%%%%%%%%%%%%%%%%%%%%%%%%%%%%%%%%
%\begin{figure}
%\begin{center}
%\resizebox{0.5\textwidth}{!}{
% \includegraphics[scale=0.5]{fig2.eps}
%}
%\caption{(a): Mobility ($\mu$) and density ($p$) vs. $T$, deduced from
%the integrated intensity and linewidth of feature A in Fig. 1.
%(b): Magnetization ($M$) vs. $T$,
%obtained with a field of 0.5 mT
%applied along the growth direction.
%}
%\label{fig2}
%\end{center}
%\end{figure}
%%%%%%%%%%%%%%%%%%%%%%%%%%%%%%%%%%%%%%%%%%%%%%%

%%%%%%%%%%%%%%%%%%%%%%%%%%%%%%%%%%%%%%%%%%%%%%
%%%%%%%%%%%%%%%%%%%%%%%%%%%%%%%%%%%%%%%%%%%%%%%

The clear observation of CR indicates that {\em at least}
a fraction of the holes are delocalized.
This is in agreement with our measurements on
low-$T_c$ films \cite{matsuda1,sanders_jap}, which showed
similar two resonance spectra althrough the resonances
were much broader and $T$-dependence was much weaker.
However, extensive earlier attempts to observe CR in GaMnAs
\cite{matsuda2} did not detect any sign of resonant absorption
within the $B$ and $\lambda$ ranges in which
both light hole (LH) and heavy hole (HH) CR in GaAs were expected.
This fact indicates that the holes in GaMnAs
are strongly localized, that the mixing of $p$- and $d$-like states makes
the effective masses of holes extremely large, or that scattering is too strong
to satisfy $\omega_c\tau > 1$.
In any case, it appears that the carriers mediating the
Mn-Mn exchange interaction are considerably
more localized in GnMnAs than in InMnAs, consistent with recent optical
conductivity \cite{hirakawa} and photoemission
experiments \cite{okabayashi}.

Feature A becomes strikingly narrow at low $T$'s.  The estimated CR mobility is
4$-$5 $\times$ 10$^3$ cm$^2$/Vs, which is one order of magnitude larger than
the low-$T$ mobilities measured by the Hall effect (see
Table I).
We speculate that
this is associated with the suppression of localized spin
fluctuations at low $T$'s.  A similar effect has been observed
in (II,Mn)VI systems \cite{brazis}.
Spin fluctuations become important when a band carrier
simultaneously interacts with a limited number of localized spins. This occurs,
for example, for magnetic polarons and for electrons in (II,Mn)IV quantum
dots. The strong in-plane localization by the magnetic field may also
result in a reduction of the number of spins which a band carrier
feels, thus increasing the role of spin fluctuations.

It is important to emphasize
that the $T$ at which the significant spectral changes
start to appear ($T_c^*$) is consistently higher than
$T_c$ in all the samples.  This fact may be
explainable in light of a recent Monte Carlo study \cite{schliemann},
which suggested that {\em short-range} magnetic order and
finite {\em local} carrier spin polarization are present for
temperatures substantially higher than $T_c$.
A more recent theoretical study \cite{elbio} explicitly predicts the existence
of $T_c^*$, corresponding to {\em clustering}.
Any such local order should
result in modifications in band structure, which in turn modify CR spectra.

%\section {Theoretical Calcuations}

\begin{figure}[b]
\begin{center}\leavevmode
\includegraphics[scale=0.6]{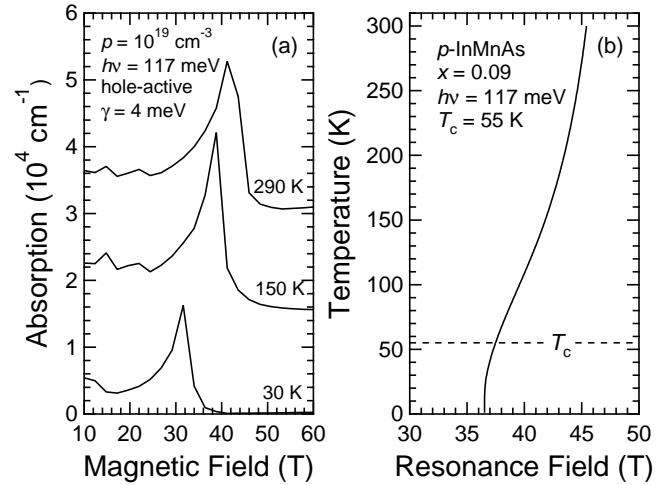}
\caption{(a) Theoretical CR spectra for sample 1 showing a shift of peak A
with temperature.  (b) Calculated temperature-dependence of the
resonance field for peak A in sample 1.
% {\bf Jun,   I would probably take out the second
%figure and just show the unbroadened results.}
}
\label{theory}
\end{center}
\end{figure}

In order to further understand the effects of ferromagnetism on band structure,
we used an 8$\times$8 {\bf k$\cdot$p} model with $s(p)$-$d$ exchange
interaction taken into account \cite{zudov,ntype}.
Each state is specified by two indices, $(n,\nu)$, where $n$ is the Landau
quantum number and $\nu$ labels the eigenvectors within each Landau
manifold.
Peak A can be identified as the HH($-$1,1) $\rightarrow$ HH(0,2)
transition \cite{sanders_jap,sanders_josc}.
We attribute the $T$-dependent peak shift to the increase
of carrier-Mn exchange interaction resulting from the increase
of magnetic ordering at low $T$.
Our calculated CR spectra are shown in
Fig. \ref{theory}(a) for bulk In$_{0.91}$Mn$_{0.09}$As with only a
minimal broadening of 4 meV.  The figure
shows the shift of peak A with decreasing $T$.  Note that the
peak in a bulk system occurrs at room $T$ at $\sim$40 T as
opposed to the heterostructures where the resonance occurs at $\sim$50 T
due to quantum confinement/strain.

%
%\begin{figure}[h]
%\begin{center}\leavevmode
%\includegraphics[scale=0.6]{shift.eps}
%\caption{Calculated $T$-dependence of the resonance field for peak A
%in sample 1.}
%\label{shift}
%\end{center}
%\end{figure}

It is easy to obtain an
exact analytical expression for this shift since it involves only the lowest
two manifolds in our model ($n = -1$, which is 1 dimensional, and $n = 0$,
which factors into two 2$\times$2 matrices for $k_z = 0$).
Furthermore, to simplify the
final expressions, we neglect the small terms arising
from the interaction with remote bands.
With these simplifications,
the cyclotron energy (at the center of the Landau subbands) has the form:
\begin{eqnarray}
E_{CR} &=& -\frac{E_g}2 + \frac14 x \langle S_z \rangle
(\alpha-\beta)  \nonumber \\
& &
+ \sqrt{\left[\frac{E_g}2 -\frac14 x \langle S_z \rangle
(\alpha-\beta)\right]^2+E_p \mu_B B} ,
\label{dE}
\end{eqnarray}
where $E_g$
is the energy gap, $E_p$ is related to the Kane momentum matrix element $P$
as $E_p=\frac{\hbar^2 P^2}{2 m_0}$, $\alpha$ and $\beta$ are $s$-$d$
and $p$-$d$ exchange constants, and $x \langle S_z \rangle$ is the
magnetization per unit cell.

In the field range of our interest ($\sim
40$ T), $\sqrt{E_p \mu_B B}$ is in the same order as $\frac{E_g}2$,
while the exchange interaction is much smaller even in the saturation
limit.  Expanding the square root in (\ref{dE}), we obtain the
final expression
\begin{equation}\label{dE1} E_{CR} \approx
\frac{E_g}2\left(\frac1{\delta}-1\right)+ \frac14 x \langle
S_z \rangle (\alpha-\beta)(1-\delta),
\end{equation}
where $\delta = E_g / (E_g^2+4 E_p \mu_B B)^{1/2}$.

If we assume that the $T$-dependence of $E_g$ and $E_p$
is small, it follows from
Eq. (\ref{dE1}) that the peak shift should follow the $T$-dependence
of $\langle S_z \rangle$.  This shift directly measures the carrier-Mn
exchange interaction.
To obtain quantitative agreement with the experiment,
one should calculate $\langle S_z \rangle$ by taking into account the
possibility of short-range ordering, as discussed above
\cite{schliemann,elbio}.
This effect could substantially modify
the band structure at low $B$. At high $B$, however, this effect
should be smoothed out by the
field-induced magnetic ordering. In the following we neglect
this effect and calculate $\langle S_z \rangle$ via standard mean-field theory
\cite{ashcroft}, solving the transcendental equation
\begin{equation}
\label{Sz}
\langle S_z \rangle
= S B_S \left(\frac{g S}{kT}\left[\mu_B B-\frac{3 k T_c \langle S_z
\rangle}{gS(S+1)}\right]\right),
\end{equation}
where $g$ is the free
electron $g$ factor, $B_S$ is the Brillouin function, and
$S=\frac52$ is the spin of the magnetic ion.

The $T$-dependence of the resonance field, calculated using
Eqs. (\ref{dE1})-(\ref{Sz}), is presented in Fig. \ref{theory}(b).
Parameters used in the calculation are $x$ = 0.09,
$T_c$ = 55 K, $E_g$ = 0.4 eV, $E_p$ = 21 eV, and $\alpha-\beta=1.5$ eV.
It shows that from room temperature to 30 K the the resonance $B$
decreases by $\sim$20\%,
approximately the result observed in the experiment.
In addition, we find that the shift is nonlinear with $T$ and
the main shift occurs at $T$'s well above $T_c$, which is also
consistent with the experiment.

We gratefully acknowledge support from NSF
DMR-0134058 (CAREER) and DARPA MDA972-00-1-0034 (SPINS).
CJS also acknowledges support from NSF DMR-9817828.
We also thank A. H. MacDonald, S. J. Allen,
and E. Dagotto for useful discussions.
 
% BibTeX users please use
%\bibliographystyle{prb}

\bigskip
\bigskip

%\bibliography{fmcr}

\end{document}